\documentclass[aps,prl,twocolumn,amsmath,amssymb]{revtex4-1}

\usepackage{amsmath,amsfonts,amssymb,bm,hyperref,xcolor,amsopn}
\usepackage[capitalize]{cleveref}
\usepackage{graphicx,subfigure,epstopdf,multirow,diagbox}
\usepackage[normalem]{ulem}
\usepackage{soul}
\usepackage{cancel}
\renewcommand\thesubsection{\arabic{subsection}}
\soulregister\cite7
\soulregister\ref7

\begin{document}

\title{Solution landscape of reaction-diffusion systems reveals a nonlinear mechanism and spatial robustness of pattern formation}

\author{Shuonan Wu$^{1}$}

\author{Bing Yu$^{1}$}

\author{Yuhai Tu$^{2}$}

\author{Lei Zhang$^{3}$}

\affiliation{$^1$School of Mathematical Sciences, Peking University, Beijing 100871, China\\ $^2$IBM T. J. Watson Research Center, Yorktown Heights, NY, USA\\ $^3$Beijing International Center for Mathematical Research, Center for Quantitative Biology, Center for Machine Learning Research, Peking University, Beijing, 100871, China.}

\begin{abstract}
Spontaneous pattern formation in homogeneous systems is ubiquitous in nature. Although Turing demonstrated that spatial patterns can emerge in reaction-diffusion (RD) systems when the homogeneous state becomes linearly unstable, it remains unclear whether the Turing mechanism is the only route for pattern formation. Here, we develop an efficient algorithm to systematically map the solution landscape to find all steady-state solutions. By applying our method to generic RD models, we find that stable spatial patterns can emerge via saddle-node bifurcations before the onset of Turing instability. Furthermore, by using a generalized action in functional space based on large deviation theory, our method is extended to evaluate stability of spatial patterns against noise. Applying this general approach in a three-species RD model, we show that though formation of Turing patterns only requires two chemical species, the third species is critical for stabilizing patterns against strong intrinsic noise in small biochemical systems.

\end{abstract}

\maketitle


\section{INTRODUCTION}

\begin{figure}[htb]
\centering
\includegraphics[width=7cm]{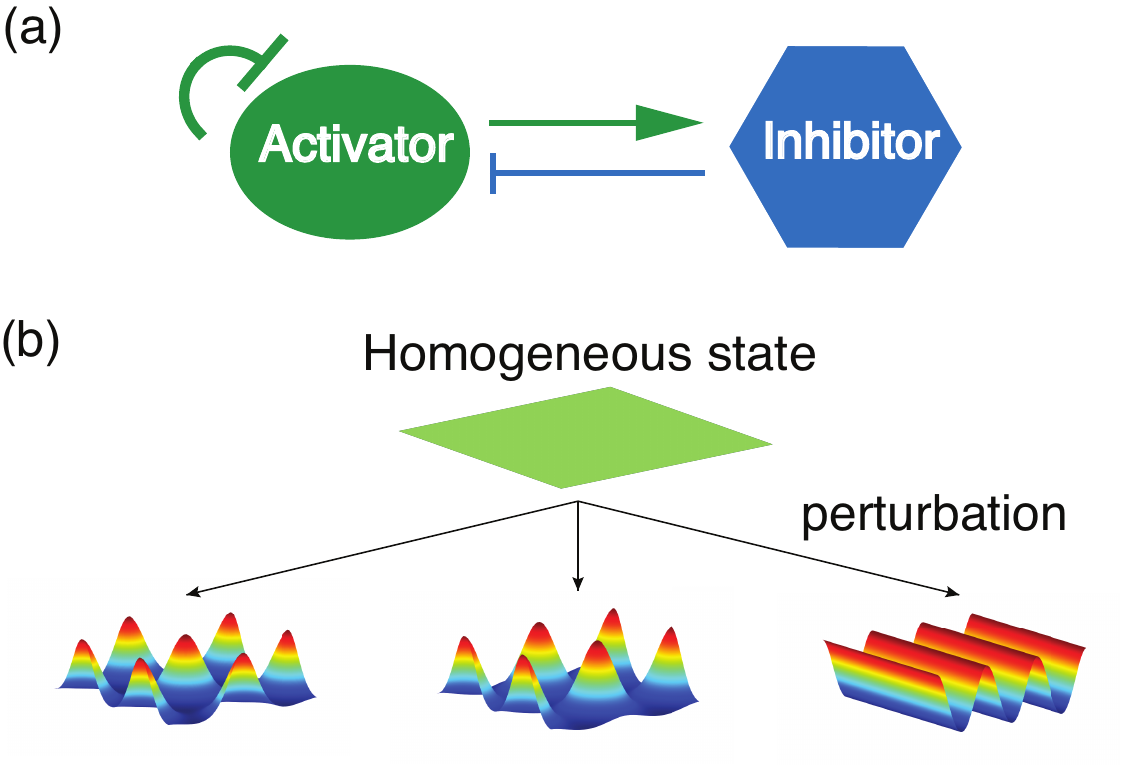}
\caption{(a) Illustration of a two-species RD system for Turing pattern; (b) Turing instability indicates small perturbation of an unstable homogeneous state can generate stable inhomogeneous states.}
\label{fig1}
\end{figure}

Alan Turing, in his seminal study on morphogenesis~\cite{1952The}, suggested that biological pattern formation can be understood by reaction-diffusion (RD) dynamics~\cite{gierer1972theory,2010Reaction,roth2011mathematics}. In the simplest scenario, nonlinear chemical reactions between two diffusive chemical species, a short-range activator and a long-range inhibitor (as illustrated in Fig.~\ref{fig1}(a)), can spontaneously generate spatial patterns~\cite{murray1982parameter,Ouyang1991, 1997Spatial,alon2006introduction}. As shown in Fig.~\ref{fig1}(b), the homogeneous (H) state with spatially uniform concentrations becomes linearly unstable against a perturbation with certain wavelength when the reaction rates and diffusion constants satisfy certain conditions leading to spatial patterns that break spatial translational symmetry of the underlying dynamics. Such mechanism is often called ``Turing instability'' or ``diffusion-driven instability''~\cite{2012Turing, 2010Reaction}.      

Many biological and physical systems can be described by RD models, for example, neuronal interactions of ocular dominance stripe formation \cite{Swindale1980A}, waves on the skin of the marine angelfish Pomacanthus  \cite{1995A}, frequency-locking phenomena in a rotating spiral wave with external periodic forcing \cite{1997Resonant}, interactions between zebrafish pigment cells \cite{nakamasu2009interactions}, trans-membrane signal transduction \cite{2004The}, mussel population density patterns \cite{2008Experimental}, etc. 


Turing instability has long been regarded as the prototype mechanism for pattern formation in homogeneous RD systems \cite{2010Reaction}. 
Most theoretical studies on Turing instability mechanism rely on linear stability analysis of the H state \cite{2013Mathematical,KOVACS2004567} and its excitable wave numbers to establish general conditions for Turing instability. However, 
there are systems that can develop stable spatial patterns independent of the Turing instability \cite{2008The,2011Pattern,smith2018beyond,2006Pattern}, 
e.g., cell polarization responsible for cellular sensing and responsiveness \cite{MORI20083684,Nandan2023.02.03.526969}.      In these cases, stable patterns are not originated from small perturbations of H state(s). However, except for Turing instability, other mechanisms of pattern formation and selection in RD systems are not well understood. 

In Turing's original paper and most of the work afterwards, noise was not considered. However, dynamics of pattern formation in realistic biological systems are subject to strong noise due to the finite number of molecules involved in biochemical reactions. Given that spatial order under noisy environment is crucial in many cellular processes~\cite{PhysRevE.79.032901,PhysRevE.80.030902}, an important question is how patterns maintain their spatial accuracy amid large spatial-temporal noise in the underlying biochemical reactions. 

To address the two aforementioned fundamental problems in RD systems, i.e., nonlinear mechanism for pattern formation and robustness of spatial patterns, we introduce a solution landscape approach to systematically search the concentration functional space for steady-state solutions. Application of the solution landscape approach to generic RD models 
reveals a general nonlinear mechanism for pattern formation in the subcritical regime before the onset of the Turing instability. 
Next, based on the solution landscape framework, we develop a method to evaluate the stability of any given spatial pattern against noise by employing the concept of ``action'' from Freidlin-Wentzell large deviation theory\cite{2012Random,LDT}. We apply the action-based stability analysis method to study a reversible three-species RD model where we show that the addition of the third species is crucial for stabilizing the spatial pattern against noise and significantly reduces fluctuations in the spatial patterns despite its weak interactions with the two pattern-formation chemical species.


\section{METHODS AND RESULTS}

\subsection*{1. The solution landscape approach for studying RD systems}
Two-species reaction-diffusion(RD) systems are simple but typical in generating spatial patterns.
The deterministic dynamics of the two-species RD system can be generally formulated as:
\begin{equation}\label{general 2-species RD}
\begin{aligned}
\frac{\partial u}{\partial t}&=D_1\Delta u +f(u,v), \\
\frac{\partial v}{\partial t}&=D_2\Delta v +g(u,v),
\end{aligned}
\end{equation}\\
where $u$ and $v$ are spatiotemporal variables representing concentrations of reactants. $D_1$ and $D_2$ are diffusion constants of two species, $u$ and $v$, respectively. Functions $f(u,v)$ and $g(u,v)$ contain nonlinear terms representing biochemical reactions.

To find all possible stationary solutions, many numerical algorithms, \emph{e.g.,} the eigenvector-following method \cite{wales1994rearrangements, doye1997surveying}, the numerical polynomial homotopy continuation method \cite{mehta2011finding}, the deflation technique \cite{farrell2015deflation}, and the WKBJ-based approach \cite{2020From}, have been developed. 
However, since these methods rely on suitable initial guesses, they may not find all the solutions and more importantly, they can not reveal the relationships between different solutions. Here, we introduce the solution landscape approach based on the generalized high-index saddle dynamics (GHiSD) to construct the solution landscapes of RD systems. 

The solution landscape is a pathway map consisting of all stationary states and their connections \cite{yin2020construction, yin2021searching}. In an energy-based (or gradient) system, the solution landscape can be efficiently constructed by the high-index saddle dynamics (HiSD) method combined with downward/upward search algorithms \cite{yin2019high,yin2020construction}, and it has been successfully applied to Ginzburg--Landau model for phase transition \cite{yin2021searching}, Landau--de Gennes model for liquid crystals \cite{yin2020construction}, the Gross--Pitaevskii model for Bose--Einstein condensation \cite{yin2022constrained}, the Lifshitz--Petrich model for quasicrystals \cite{Yin2021transition}, etc. 
In this paper, we introduce GHiSD, which is modified from HiSD, in order to construct the solution landscapes for non-gradient systems.

Mathematically, we view stable stationary states as the sinks of the dynamic systems, and unstable stationary states are the saddle points. Here we adopt the index theory from infinite dimensional Morse theory\cite{Chang1993} to give a rigorous mathematical description of topological invariants for saddle points of partial differential equation systems. The Morse index characterizes the nature of the nondegenerate saddle point, and an index$-k$ saddle point ($k$-saddle) is a stationary state whose Jacobian matrix has exactly $k$ eigenvalues with a positive real part \cite{1989Morse}. From this perspective, a sink ( referring to the fixed-point type in this paper) can be regarded as a 0-saddle point and two neighboring sinks are connected by a 1-saddle point (i.e., transition state). Every stationary solution is assigned an index. Particularly, a homogeneous solution with positive index indicates the occurrence of Turing instability.

In Eq.~\ref{general 2-species RD}, let $\boldsymbol{u}=(u,v)^\top$, $\boldsymbol{F}(\boldsymbol{u})=(D_1\Delta u +f(u,v),D_2\Delta v +g(u,v))^\top$ and this RD system could be written as: $\dfrac{\partial{ \boldsymbol{u}}}{\partial t}=\boldsymbol{F}(\boldsymbol{u})$, 
the formulation of GHiSD for searching a $k$-saddle ($k$-GHiSD) can be written as follows.
\begin{equation}\label{flow}
\left\{\begin{array}{l}
\dfrac{\partial{ \boldsymbol{u}}}{\partial t}=\left(I-2 \sum\limits_{j=1}^{k} \boldsymbol{w}_{j} \boldsymbol{w}_{j}^{\top}\right) \boldsymbol{F}(\boldsymbol{u}), \\
\dfrac{\partial{ {\boldsymbol{w}}_{i}}}{\partial t}=\left(I-\boldsymbol{w}_{i} \boldsymbol{w}_{i}^{\top}\right) J \boldsymbol{w}_{i}-\sum\limits_{j=1}^{i-1} \boldsymbol{w}_{j} \boldsymbol{w}_{j}^{\top}\left(J+J^{\top}\right) \boldsymbol{w}_{i}, \\
\qquad i=1, \cdots, k.
\end{array}\right.
\end{equation}

The $k$-GHiSD involves a spatiotemporal variable $\boldsymbol{u}$ representing the concentrations of various substances, and $k$ direction variables $\{ {\boldsymbol{w}}_{i} \}_{i=1}^k$ that approximate an orthonormal basis of the maximum subspace (unstable subspace) of the $k$-saddle. $\boldsymbol{F}$ is an operator mapping spatiotemporal concentration function $\boldsymbol{u}$ to function $\boldsymbol{F}(\boldsymbol{u})$. $J=J(\boldsymbol{u})$ is the Jacobi operator of $\boldsymbol{F}(\boldsymbol{u})$.

The downward search algorithm is to apply GHiSD starting from a high $m$-saddle  $\boldsymbol{u}^*$  as a parent state to search low $k$-saddle ($k<m$). 
The initial searching position $\boldsymbol{u}_0=\boldsymbol{u}^* \pm \delta \boldsymbol{u}$ is chosen to push the system away from $\boldsymbol{u}^*$, and the pushing direction $\delta \boldsymbol{u}$ is along a linear combination of $(m-k)$ vectors whose negative eigenvalues have the smallest magnitudes chosen from $\{ {\boldsymbol{w}}^*_{i} \}_{i=1}^m$, the eigenvectors of Jacobi matrix at $\boldsymbol{u}^*$. 

Also we have upward search algorithm as an auxiliary to find higher index saddles. The combination of GHiSD and downward/upward search navigates the entire search up and down to construct the complete solution landscape. The details of the GHiSD algorithm and downward/upward search in constructing the solution landscape are documented in Appendix A.

\subsection*{2. Pattern formation in subcritical Turing regime revealed by the solution landscape approach} 

Although our methodology applies to RD systems in general, for simplicity we will mainly focus on the 2-D Schnakenburg model \cite{SCHNAKENBERG1979389}, a minimal chemically realistic model that can give rise to Turing patterns \cite{2012Hopf,2020Spatial,2020From}.  The Schnakenburg model is described by PDEs in $ \Omega=[0,1] \times [0,1]:$

\begin{equation}\label{Schnakenburg}
\begin{aligned}
\frac{\partial u}{\partial t}&=\Delta u +\eta\left(a-u+u^{2} v\right), \\
\frac{\partial v}{\partial t}&=d \Delta v +\eta\left(b-u^{2} v\right).
\end{aligned}
\end{equation}\\
The model corresponds to the following reactions \\
$$
2u+v \xrightarrow{\eta} 3u,\quad \phi \xrightarrow{\eta a} u,\quad u \xrightarrow{\eta} \phi, \quad \phi \xrightarrow{\eta b} v.
$$
Here $u,v$ represent the density of activator and inhibitor respectively. Parameter $d$ denotes the relative diffusion constant of two species and $\eta$ represents a relative balance between the diffusion and the chemical reaction. Parameter $a$ and $b$ are the constant speeds of two species produced in the domain uniformly respectively. The boundary condition is generally set as no-flux condition for both species, i.e. $\frac{\partial u}{\partial n}|_{\partial \Omega}=\frac{\partial v}{\partial n}|_{\partial \Omega}=0 $. 

Many theoretical analyses have been done based on local linearization and excitable modes \cite{SARFARAZ201733} and have explained the influence of domain size of this model to generate Turing instability \cite{Anotida,articledomain}. Here, we use the solution landscape approach to map out all the steady state solutions in the parameter space.

We start with the unique, homogeneous stationary solution, i.e., H state, $(u_0,v_0)=(a+b,\frac{b}{(a+b)^2})$ in the Schnakenburg model (Eq.~\ref{Schnakenburg}). 
Using the linear stability analysis \cite{2013Mathematical} (see details in Appendix B), the Morse index of H state can be explicitly calculated and are shown in Table~\ref{table1}.

\begin{table}[hbt]
\caption{\footnotesize Morse index of the H state with different $d$ in the Schnakenburg model ($\eta=200, a=\frac{1}{3}, b=\frac{2}{3} $).}\label{table1}
\setlength\tabcolsep{1.7pt}
\renewcommand\arraystretch{1.1}
\centering
\begin{tabular} {cccccc} 
\hline \hline 
$d$ & $\left(0,44.6\right)$ & $\left(44.6,47.4\right)$ & $\left(47.4,58.4\right)$ & $\left(58.4,74.9\right)$ & $\left(74.9, +\infty \right)$\\
\hline 
Index & 0 & 2 & 3 & 5 & 7\\
\hline \hline
\end{tabular}
\end{table}

The H state is a stable sink when $d$ is less than a critical value $d_0$: $d\le d_0=44.6$. When $d>d_0$, the H state losses its linear stability and stable spatial solutions emerge from the H state via pitchfork bifurcations, which is exactly the Turing mechanism. Two questions immediately present themselves: How many stable solutions are there in the supercritical regime ($d>d_0$)? More importantly, are there any stable solutions in the subcritical regime ($d<d_0$)? We address these two questions below by using the solution landscape approach.   

\begin{figure*}[thb]
\centering
\includegraphics[width=\linewidth]{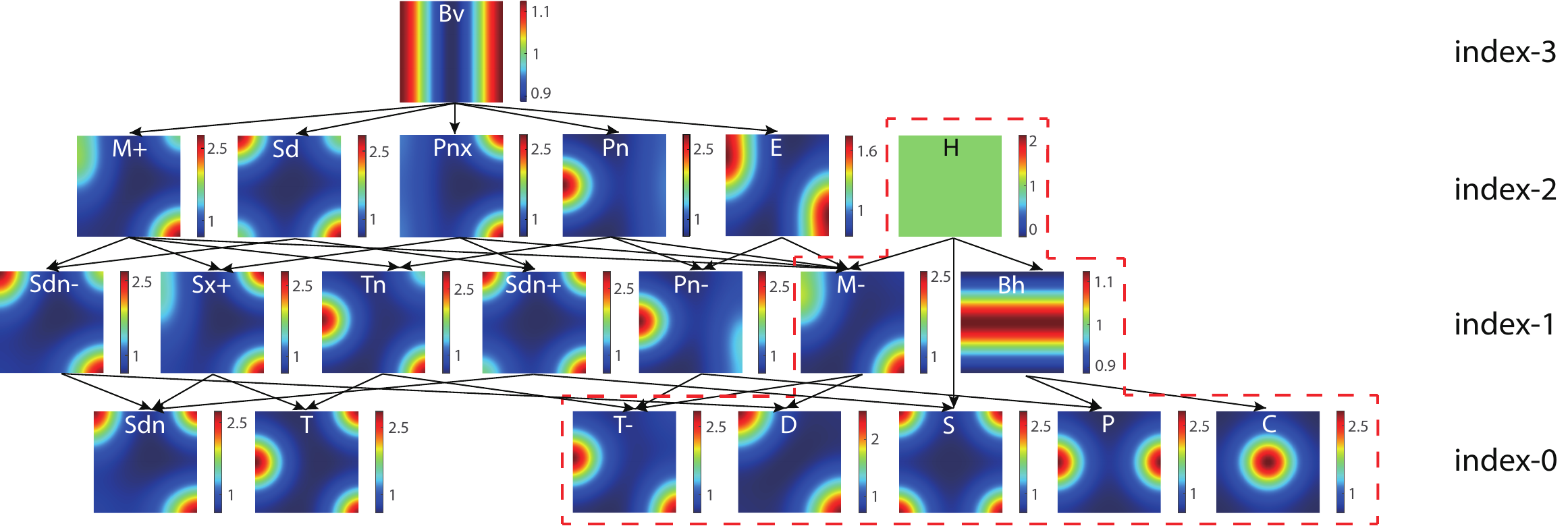}
\caption{ Solution landscape in the supercritical regime ($d=46>d_0$) . Patterns of $u$ and $v$ are very alike so only $u$ is shown in the figure ( the same way in the following figures).
Patterns induced by Turing instability (use the index-2 H as the parent state) is highlighted in red dashed line area. 
Using the index-3 Bv as the parent state, the solution landscape shows more saddle points and new stable sinks.}
\label{fig2}
\end{figure*}

\textit{Pattern formation in the supercritical regime ($d>d_0$).} 
To find solutions in the supercritical regime, 
we start with the H state as the parent state. Here, we choose $d=46$ where the H state is a 2-saddle in the solution landscape. By applying the downward search algorithm, we can find all possible stationary states (1-saddles and stable sinks) that originate from the H state as shown in the region enclosed by the red dotted line in Fig.~\ref{fig2}. 
The H state gives rise directly to one sink (S) and two 1-saddles M- and Bh, which subsequently give rise to T- $\&$ D sinks and P $\&$ C sinks, respectively. 
Note that the S, P and C states correspond to the same sink due to the symmetry of the pattern and boundary condition. 
We validate this result by direct simulations of Eq.~\ref{Schnakenburg} using random perturbations of the H state (see details in Appendix C). 

Although multiple sinks are generated from the H-state through Turing instability, it is unclear whether it is the only way to generate stable solutions. To address this question, we apply the upward search starting from the known sinks to find other possible parent states. 
In particular, we pick T- as the starting state and use the upward search to find an upward pathway T- $\rightarrow$ Tn $\rightarrow$ Pn $\rightarrow$ Bv, in which Bv is a band-shape, 3-saddle that has a higher index than that of H.
Using Bv as the new parent state, we are able to compute a complete solution landscape shown in Fig.~\ref{fig2}. It shows that there exist more 2-saddles (e.g. M+, Sd, Pnx, Pn and E) than H, and they connect new 1-saddles and two new sinks, Sdn and T.
Both these new sinks are easily found by using downward search through downward pathways from Bv, for example, Bv $\rightarrow$ M+ $\rightarrow$ Sx+ $\rightarrow$ Sdn and Bv $\rightarrow$ Pn $\rightarrow$ Tn $\rightarrow$ T, but there are no pathway connecting H to Sdn and T,  indicating that neither of these stable states (Sdn and T) can be generated by the Turing instability.

\begin{figure*}[thb]
\includegraphics[width=\linewidth]{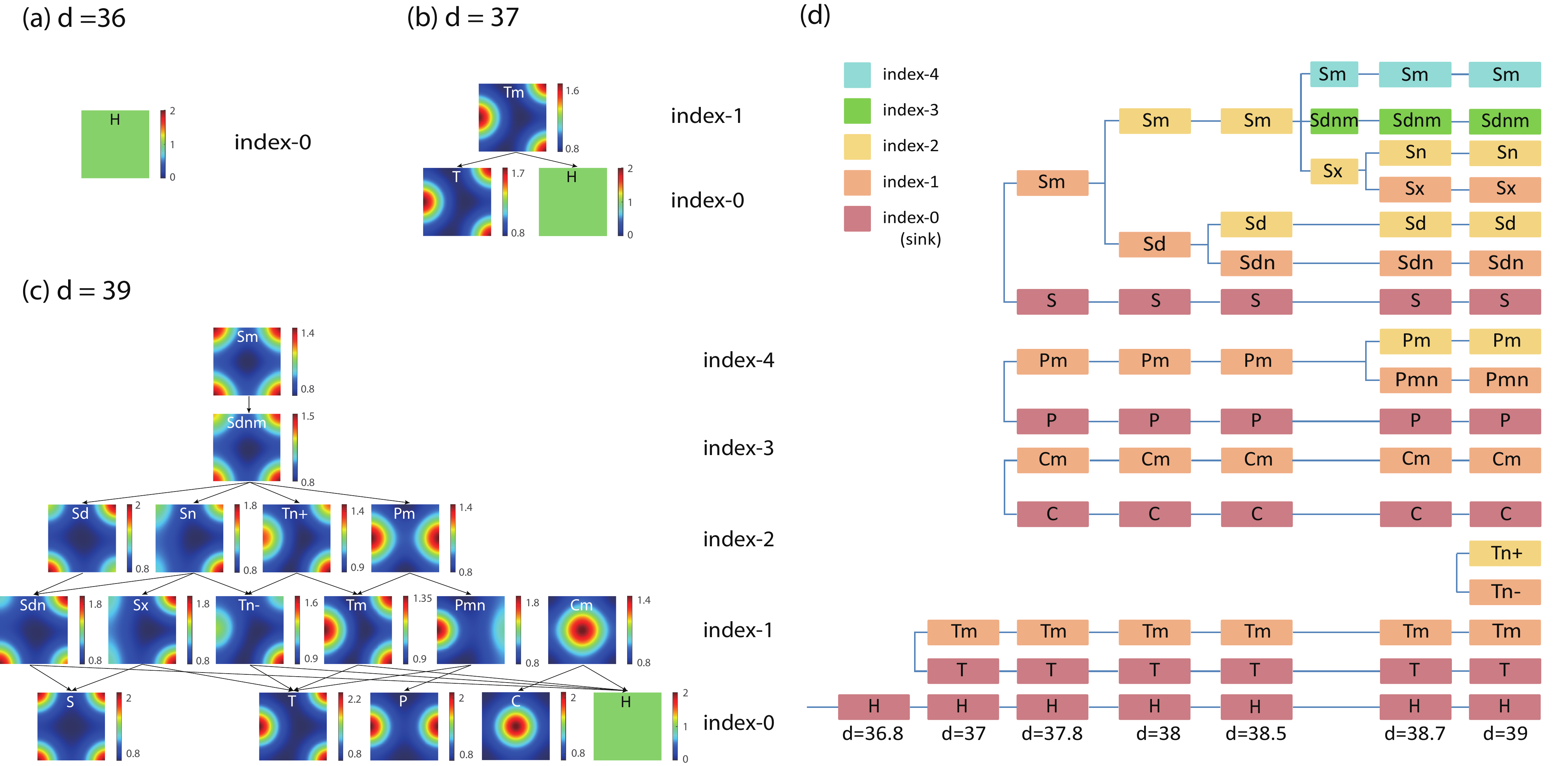}
\caption{Solution landscapes in the subcritical regime ($d< d_0=44.6$). (a)-(c) Solution landscapes when relative diffusion speed $d=36,37,39$, respectively. (d) Bifurcation diagram by tracing the solution landscapes with changing $d$.
}
\label{fig3}
\end{figure*}


\textit{Pattern formation in the subcritical regime ($d<d_0$).}   The appearance of Bv, whose Morse index is higher than H, suggests that there may exist other mechanisms besides Turing instability for pattern formation. This prompt us to investigate the solution landscape of the Schnakenburg model in the subcritical regime ($d< d_0=44.6$) where the H state is stable.
We find that the stable H state is the only stationary state for $d\leq 36$ (Fig.~\ref{fig3}(a)). As $d$ increases to $37$, a pair of stationary states, Tm (1-saddle) and T (stable sink), emerges via a saddle-node bifurcation (Fig.~\ref{fig3}(b)). If $d$ becomes larger, more stationary states can be identified by applying the GHiSD method. For example, when $d=39$, Sm is found to be a 4-saddle. Using Sm as the parent state, we construct the corresponding solution landscape in Fig.~\ref{fig3}(c).  It shows that four inhomogeneous sinks (S, T, P and C) exist besides H. Although S, P and C can be generated by Turing instability in the supercritical regime when H loses its stability ($d=46$ in Fig.~\ref{fig2}), the emergence of these three states in the subcritical regime are due to some other mechanism. By applying GHiSD  for different values of $d$ in the subcritical regime,  we obtain the bifurcation diagram (Fig.~\ref{fig3}(d)), which clearly shows that there exist a series of saddle-node bifurcations that generates stable sinks (T, S, P and C) while H remains stable.  
Note that the stable sinks can arise by infinitesimal perturbation from the H state in the supercritical regime. In contrast, when H is a stable sink in the subcritical regime, even though there exist transition pathways connecting H to certain inhomogeneous stable sinks such as T, S and C, a finite perturbation is required to overcome the transition barrier determined by transitional 1-saddle points, namely Tm, Tn-, Sdn and Cm.

In  Supplemental Material Fig.~1 \cite{SM} we compare the basins of attraction of sinks when $d=39$. Inhomogeneous sinks have quite large basins of attraction, indicating that these sinks are non-negligible subcritical patterns before the onset of the Turing instability. 

\begin{figure}[htb]
\centering
\includegraphics[width=8cm]{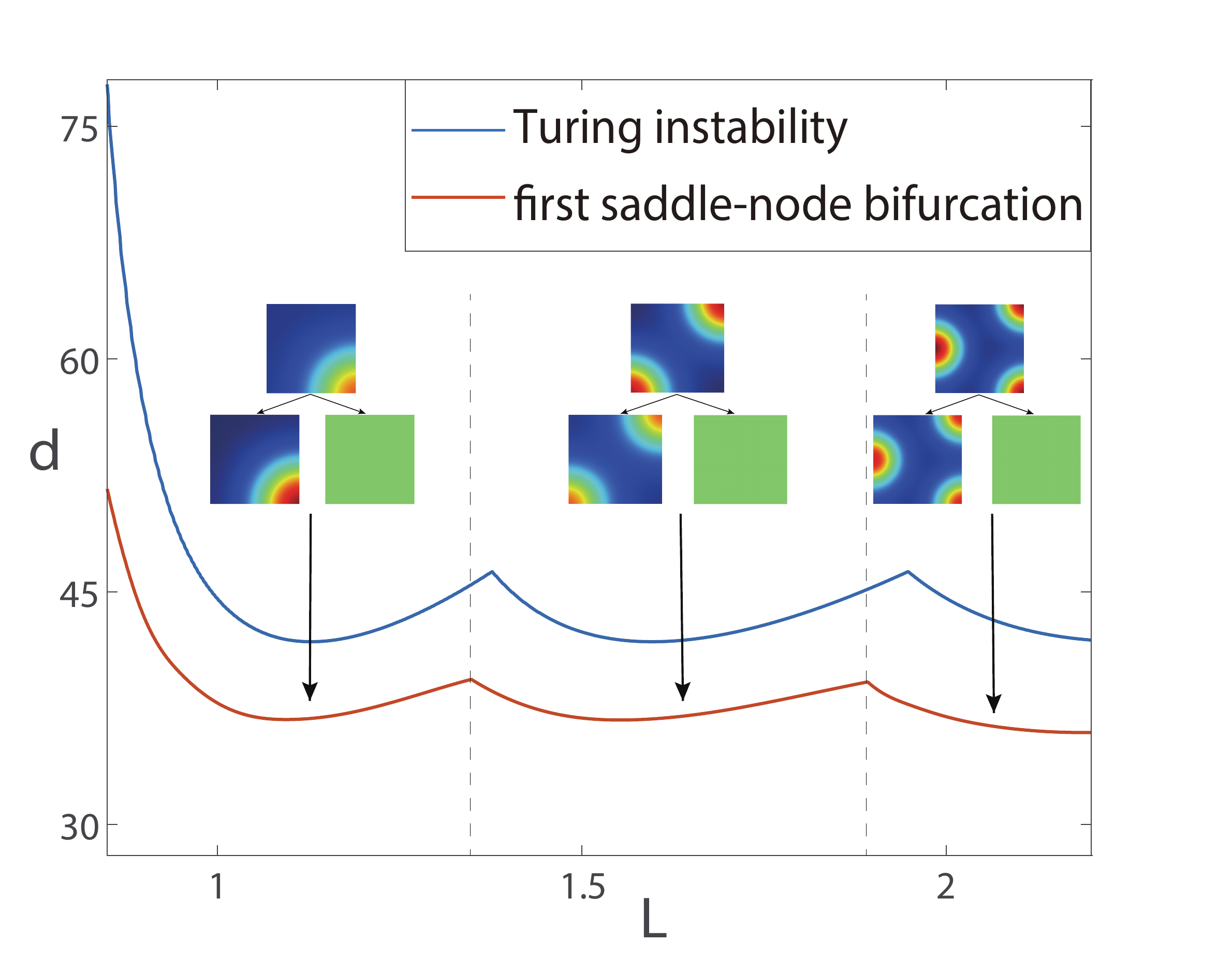}
\caption{
Phase diagram for different $d$ and square domain sizes $L$ with $\eta=50$ fixed. The blue curve represents the Turing instability. The red curve represents the first saddle-node bifurcation, including three different saddle-node bifurcations (corresponding solution landscapes are inserted) separated by black dash lines.
}\label{fig4}
\end{figure}

To check whether the subcritical saddle-node bifuraction mechanism depends on the system size, we also study the Schnakenburg model for different $L$ (size of the square domain) and $d$. The phase diagram in the $(L,d)$ space with $\eta=50$ fixed is shown in Fig.~\ref{fig4} where we plot the first saddle-node bifurcations with the corresponding solution landscapes as well as the Turing instability onsets for different values of $L$.
It is clear from Fig.~\ref{fig4} that the first saddle-node bifurcation always occurs before Turing instability independent of system size. This nonlinear mechanism greatly increases the parameter space where stable patterns exist beyond those originated from the Turing instability. We have also checked different boundary conditions do not alter the main conclusions of this paper (see  Supplemental Material  Fig.~2  \cite{SM} for results with periodic boundary condition). 


Finally, to test the universality of our findings in the Schnakenburg model, we examine the solution landscape of the Gierer-Meinhardt model \cite{gierer1972theory, RevModPhys.66.1481}. The results shown in  Supplemental Material Fig.~3 \cite{SM} support the existence of a general subcritical saddle-node bifurcation mechanism for pattern formation in RD systems.

\subsection*{3. Stability of spatial patterns against noise in solution landscape}

The precise positioning of biomolecules is essential for many biological processes such as cell division and development. However, for biological systems, the size of the system and the number of molecules in the system are relatively small. As a result, dynamics underlying pattern formation in biological systems are subject to finite size effects and large stochastic fluctuations (noise). 
Therefore, understanding stability of various pattern forming states against noise is critical for understanding robustness of precise positioning in biological systems. For example, although a two-species RD system can generate spatial patterns, Murray \& Sourjik recently highlighted the significance of an additional third species for enhancing the positional accuracy of Turing patterns in noisy biological systems \cite{murray2017self}. 

Here we consider the reversible three-species reaction–diffusion model proposed by Zhang et al \cite{zhang2023free} where it was shown numerically that 
patterns formed in a 2-species model are extremely sensitive to noise, and addition of the third specie in the RD system can stabilize the patterns even with weak interaction strength (or equivalently small reaction rates). 
Reactions of this reversible model involve three linear reactions between each pair of three species $X_1,X_2,X_3$ and a nonlinear auto-catalytic reaction between $X_1$ and $X_2$:
\begin{equation*}
   X_1 \underset{k_{21}}{\stackrel{k_{12}}{\rightleftharpoons}} X_2 \text{, } X_2 \underset{k_{32}}{\stackrel{k_{23}}{\rightleftharpoons}} X_3 \text{, } X_3 \underset{k_{13}}{\stackrel{k_{31}}{\rightleftharpoons}} X_1 \text{, } X_1+2 X_2 \underset{\tilde{k}_{21}}{\stackrel{\tilde{k}_{12}}{\rightleftharpoons}} 3 X_2. 
\end{equation*}

In the absence of noise, dynamics of three species' concentration $u_1,u_2,u_3$ can be described by deterministic PDEs:
\begin{equation}\label{3-species}
\begin{aligned}
\frac{\partial u_1}{\partial t} = & -\tilde{k}_{12} u_1 u_2^2+\tilde{k}_{21} u_2^3+k_{21} u_2-k_{12} u_1 \\
& -k_{13} u_1+k_{31} u_3+D_1 \frac{\partial^2 u_1}{\partial x^2}, \\
\frac{\partial u_2}{\partial t} = & \tilde{k}_{12} u_1 u_2^2-\tilde{k}_{21} u_2^3-k_{21} u_2+k_{12} u_1 \\
& +k_{32} u_3-k_{23} u_2+D_2 \frac{\partial^2 u_2}{\partial x^2}, \\
\frac{\partial u_3}{\partial t} = & -\left(k_{32}+k_{31}\right) u_3+k_{13} u_1+k_{23} u_2\\
&+D_3 \frac{\partial^2 u_3}{\partial x^2},
\end{aligned}   
\end{equation}
where $x \in \Omega = (0,6)$ and $u_i(i=1,2,3)$ satisfy zero-flux boundary condition. The reaction rate constants $\tilde{k}_{12}, \tilde{k}_{21}, k_{12}, k_{21}$ between $X_1$ and $X_2$ are fixed: $ \tilde{k}_{12}=1.67 \times 10^{-5},$$\tilde{k}_{21}=2.40\times 10^{-6},$$k_{12}=0.5,k_{21}=3.6$. The diffusion coefficients are chosen as $D_1=D_3=1.8$, $D_2=0.012$ such that the diffusion ratio $d=D_1/D_2=150$ is deep inside the Turing instability regime. 

The reactions between the additional third species $X_3$ and $X_1,X_2$ are described by four reaction rates $k_{13}=k_{23}=0.0139\tau,k_{31}=0.0416\tau,k_{32}=0.00139\tau$ where we introduce a scaling constant $\tau$ to control the interaction strength between $X_3$ and $X_{1,2}$ with $\tau=0$ representing the 2-species model. Note that the interaction is weak even when $\tau=1$ given the relatively small rates involving $X_3$. Our goal is to determine the stable solutions (sinks) and their stability as $\tau$ varies by using the solution landscape approach.   

\begin{figure*}[!tb]
\includegraphics[width=\linewidth]{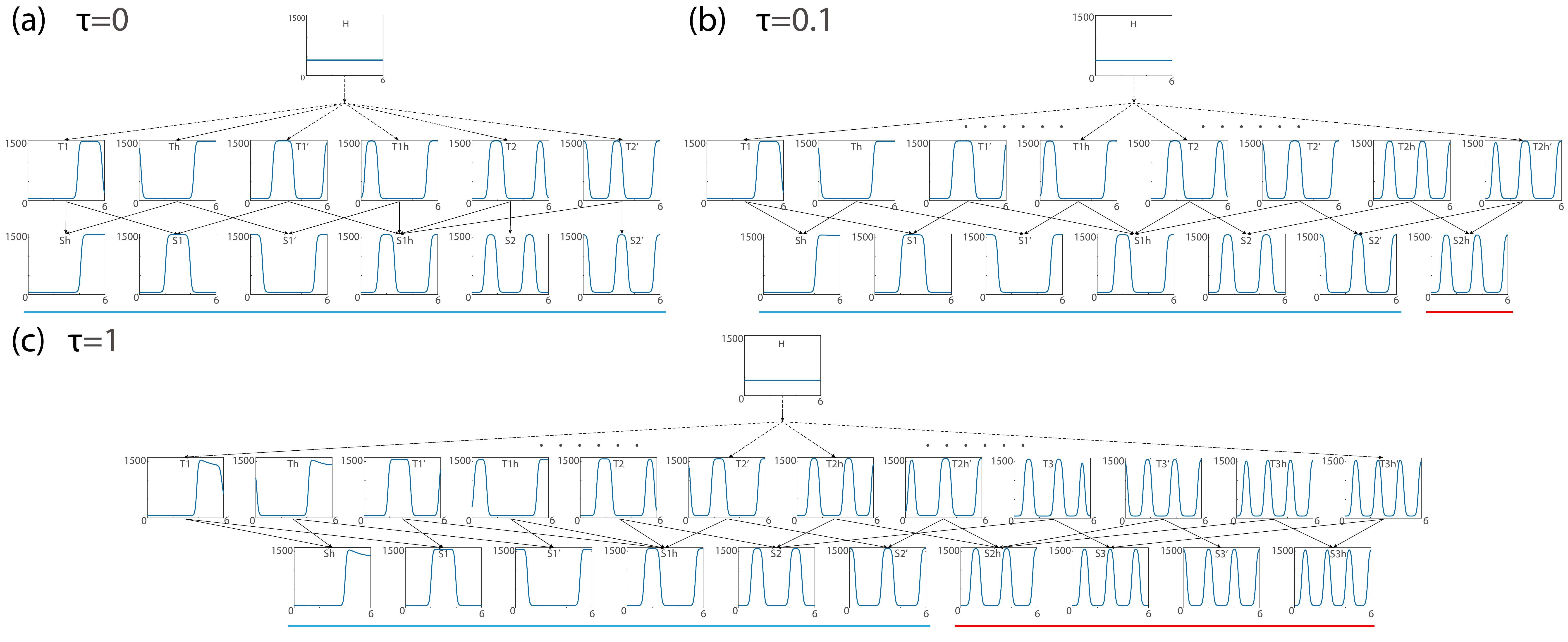}
\caption{
Solution landscapes under parameter $\tau=0, 0.1, 1$. Curves stand for spatial concentration of $X_2$ in the region $\Omega = (0,6)$. Sinks, 1-saddles and transition pathways are shown while high index saddles are omitted. ``S'' in the name of solutions represents for sinks and ``T'' for 1-saddles. The homogeneous state H is a 24-saddle in each solution landscape.  
}\label{fig5}
\end{figure*}

\textit{$X_3$ introduces more stable patterns with higher wave numbers.} The solution landscapes corresponding to different values of $\tau(=0,0.1,1)$ are constructed. In Fig.~\ref{fig5}, we show all 1-saddles and sinks as well as transition pathways connecting them.
In the two-species model (i.e. $\tau =0$), the solution landscape (Fig.~\ref{fig5}(a)) has six sinks (
 underlined with a blue line) and six 1-saddles with relatively smaller wave numbers. As $\tau$ increases, more stable modes emerges in addition to the existing stable modes in the 2-species model. These additional stable modes (1 for $\tau=0.1$, 4 for $\tau=1$, underlined with a red line respectively) have higher wave numbers as shown in Fig.~\ref{fig5}(b)\&(c). 
The bifurcation diagram containing sinks and 1-saddles are documented in the Supplemental Material  Fig.~4 \cite{SM}.

The emergence of these new patterns as $\tau$ increases can be understood in Supplemental Material  Fig.~4(a) \cite{SM}, which shows the real part of maximal eigenvalue ($max (Re \lambda)$) of 1-saddles (dotted line) and sinks (solid line) when $\tau$ increases from $0$ to $1$. Patterns with more than two peaks emerge via bifurcations when $\tau$ reaches some positive critical values when their corresponding $max (Re \lambda)$ becomes positive. Three consecutive bifurcations highlighted by red circles in Supplemental Material  Fig.~4(a) \cite{SM} are shown in details in Supplemental Material  Fig.~4(b)-(d) \cite{SM} near their bifurcation points where newly generated stable stripe-patterns are also plotted in insets. Wave numbers of newly generated patterns increase $\frac{1}{2}$ for each consecutive bifurcation as $\tau$ is increased.

\textit{$X_3$ enhances stability of the patterns.} 
Besides creating more stable patterns with higher wave numbers, the most important effect of the third species is to stabilize the 2-species solutions ($\tau=0$). As shown in Supplemental Material Fig.~4(a) \cite{SM}, as $\tau$ approaches 0, the maximum eigenvalues for the stable solutions are extremely small indicating %
poor positioning accuracy of the patterns in the presence of noise. 
However, the introduction of $X_3$ alters the stability of sinks as the negative maximal eigenvalue increases significantly with $\tau$. 

\begin{figure}[!tb]
\includegraphics[width=\linewidth]{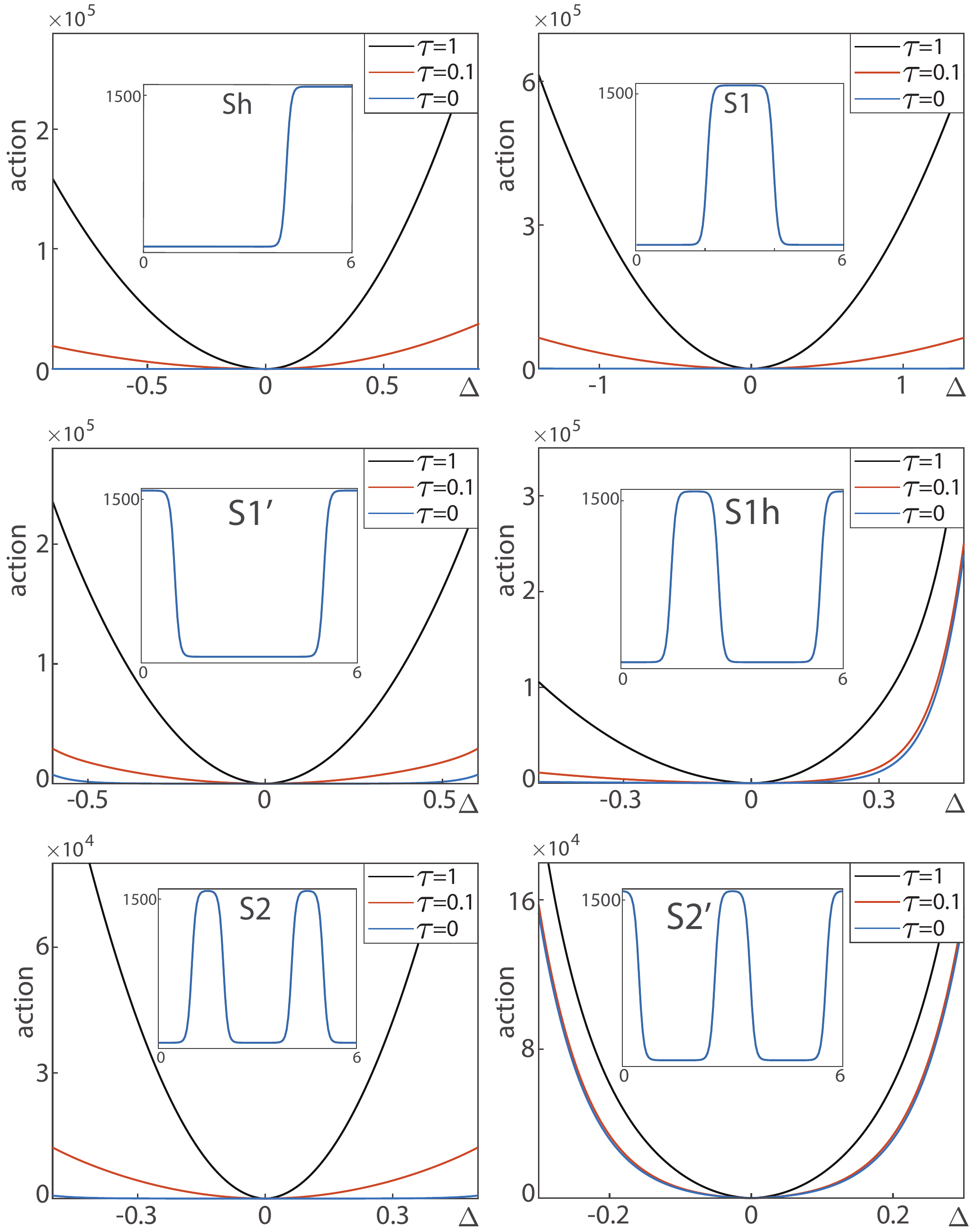}
\caption{
The action-deviation ($\Delta$) relations with $\tau = 0, 0.1, 1$ for the 6 sinks emerged in the 2-species model (see Fig.~\ref{fig5}(a)). The solution profiles for $X_2$: $u_2(x)$  are shown in the insets. 
}\label{fig6}
\end{figure}

To quantify the stability of the spatial patterns, we employ ``action'' based on Freidlin-Wentzell large deviation theory within the framework of solution landscape. In particular, the action of a given path $\varphi:[0,1] \rightarrow \mathcal{H}$ from  $\varphi(0)$ to $\varphi(1)$ is defined as follows~\cite{heymann2008geometric}:
\begin{equation}\label{action}
    \hat{S}(\varphi)=\inf _{T>0} \inf _{\psi \in \bar{C}_{\varphi}(0, T)} S_T(\psi),
\end{equation}
where $\bar{C}_{\varphi}(0, T)$ stands for all absolutely continuous functions on $[0,T]$ whose image in the Hilbert space $\mathcal{H}$ is the path $\varphi([0,1])$. Action $\hat{S}(\varphi)$ is parametrization free. It can be regarded as the cost of changing the system from state $\varphi(0)$ to state $\varphi(1)$ along the path $\varphi$, and plays the role of energy cost in non-gradient systems.
$S_T(\psi)$ is determined by the Lagrangian of the dynamics Eq.~\ref{3-species} (see details of the model's action formula in Appendix D):
\begin{equation}\label{actionInt}
    S_T(\psi)= \begin{cases}\int_0^T L(\psi, \dot{\psi}) \mathrm{d} t & \text {if } \psi \text { is absolutely continuous } \\  
    &\text { and the integral converges, } \\ 
    +\infty & \text { otherwise, }\end{cases}
\end{equation}

Since pattern formation results in breaking of translation invariance in the spatially homogeneous system, the translation degree of freedom reflects the (approximate) degeneracy of solutions, and it corresponds to the least costly mode (the ``soft" mode) in the system. Therefore, to evaluate the stability of the spatial patterns, we take the deviation $\Delta$ to be the spatial translation distance and compute the action $S(\Delta)$ along the path of deviation $\Delta$ numerically (see Appendix D). 

In Fig.~\ref{fig6}, the action-deviation ($S-\Delta$) relations for different values of $\tau = 0, 0.1, 1$ are shown for all six stable solutions ( Sh, S1, S1', S1h, S2 and S2') that exist in the 2-species model ($\tau=0$). It is clear that the action curve of each sink bends up as $\tau$ increases indicating the stabilizing effect of $X_3$. The $S-\Delta$ relations of the 6 sinks at $\tau=0$ are shown in detail in Supplemental Material  Fig.~5(a) \cite{SM}. With the exception of S2', all the sinks show weak resistance to small shift indicating their weak stability against noise. In particular, Sh and S1 can shift freely with near zero action cost in a rather wide range of $\Delta$, which implies that each of them is degenerate and lies in a one-dimensional solution manifold of Eq.~\ref{3-species}. 


To quantify the effect of $X_3$, we take the rightward shift (i.e. $\Delta \geq 0$) as an illustration (since $S-\Delta$ curves for Sh, S1h are not axisymmetric, the shift towards two directions are different). As shown in Supplemental Material  Fig.~5(b) \cite{SM}, the action-deviation relations under different values of $\tau$ in log-log coordinates imply that $S$ depends on $\Delta$ quadratically with its coefficient $C(\tau)$ depending on $\tau$:


\begin{equation} \label{C-tau}
S(\Delta)=C(\tau) \Delta^2. 
\end{equation}
To the leading order, we have:
\begin{equation} \label{expan}
C(\tau)=\epsilon+\eta \tau + \mathcal O (\tau^2),    
\end{equation}
where $\epsilon$ and $\eta$ are different for different solutions.  Table~\ref{table2} shows $\epsilon$ and $\eta$ for the six solutions that exist in the 2-species model.

\begin{table}[hbt]
\caption{\footnotesize Expansion coefficient $\epsilon$ and $\eta$ for six stable solutions 
}\label{table2}
\setlength\tabcolsep{4pt}
\renewcommand\arraystretch{1.2}
\centering
\begin{tabular} {ccccccc} 
\hline \hline 
\text{ } & Sh & S1 & S1' & S1h & S2 & S2' \\
\hline
$\epsilon$ & {$<10^{-4}$} & {$<10^{-4}$} & $170$ & $2.4 \times 10^4$ &  $60$  &  $2.7 \times 10^5$  \\
\hline 
$\eta (\times 10^5)$ & $4.7$ & $3.1$ & $5.9$ & $3.1$ &  $2.3$  & $3.8$   \\
\hline \hline
\end{tabular}
\end{table}

The value of $\epsilon$ reflects the resistance to positional fluctuations when $\tau=0$, and $\eta$ characterizes the growth of such resistance as $\tau$ increases (larger $C$ leads to smaller fluctuation and thus higher accuracy). Given that $\tau \sim \mathcal O(1)$ has already been regarded as weak coupling for $X_3$ in the system, the large value of the ratio $\dfrac{\eta}{\epsilon}$ reflects the strong effect of $X_3$ in stabilizing the sink solutions. The quadratic action gives rise to an effective Ornstein-Uhlenbeck process to describe the time evolution of shift $\Delta$ under noise:
\begin{equation*}
d \Delta_t = -2(\epsilon+\eta \tau)\Delta_t dt + \sigma dW_t  . 
\end{equation*}
Thus  $\Delta_t \rightarrow \mathcal N \left(0 , \dfrac{\sigma^2}{4(\epsilon+\eta \tau)} \right)$ as $t \rightarrow +\infty$, which clearly shows that the introduction of the third species ($\tau>0$) increases the stability of the sink solutions. Without $X_3$, stability of the sinks is controlled by $\epsilon$, which depends on the curvature of concentration functions near the boundary (see full derivation in Appendix D). Therefore, for solutions Sh,S1,S1' and S2 where $\epsilon$ is small due to their approximate translation invariance near the boundary, their stability is provided  predominantly by $X_3$. Even for solutions S1h and S2' with relatively larger $\epsilon$, their stability is enhanced significantly by $X_3$ given the large value of $\eta$ (see Table~\ref{table2}).   

\section{DISCUSSION}
In summary, 
we have developed an efficient numerical method to systematically construct the solution landscape of RD systems allowing us to 
find all possible stationary states and identify their relationships. By applying this general method to prototype RD models such as the Schnakenburg model and the Gierer-Meinhardt model, we discover a general nonlinear pattern formation mechanism wherein multiple stable heterogeneous states can emerge via saddle-node bifurcations and co-exist with the homogeneous state in the subcritical Turing regime before the onset of the linear Turing instability. 

Furthermore, within the solution landscape framework, we developed a method to evaluate the stability of spatial patterns quantitatively by using action to describe the cost of deviation from the pattern caused by noise. To demonstrate its utility, this general method is applied to explain robust positioning of spatial patterns in a reversible three-species RD system with large intrinsic noise. Our method reveals the underlying mechanism for robust self-positioning even for very weak interactions of the third species with the other two chemical species in the system.


As a generalization of HiSD developed to construct the solution landscape of gradient systems, the GHiSD method presented in this study can be applied in generic PDEs including non-gradient systems to search for multiple stationary solutions and to uncover the pathways that connect these solutions. Besides the RD systems studied in this paper, GHiSD could also be used in convection-diffusion equations \cite{1980Numerical, 1983High}, transport equations \cite{1990Introduction} and nonlinear Schr\"{o}dinger Equations \cite{2009The}. However, for a model containing discrete variables (such as the Boolean type variable), HiSD and GHiSD fail to solve the minima and saddle points on the discrete energy landscape, because the execution of iteration in saddle dynamics depends on differentiation and the information of Hessian matrix. 


From the theoretical and computational perspective, the solution landscape approach provides an efficient tool and a unifying view of pattern formation in RD systems. 
The general solution landscape approach could be instructive in studying pattern selection in synthetic systems \cite{2014Cooperativity}, designing programmable reaction-diffusion systems\cite{RAJASEKARAN2024345}, as well as understanding network robustness \cite{scholes2019comprehensive}, from micro-scale systems such as cell regulation to macro-scale systems such as ecological networks \cite{KUMAR2021100948}.

\section*{ACKNOWLEDGEMENTS}
We would like to thank Prof. Ou-yang Qi for fruitful discussions. L. Z. is supported by the National Natural Science Foundation of China (No.12225102, T2321001, and 12288101) and the National Key Research and Development Program of China  2021YFF1200500. Y. T. is supported by a NIH grant (R35GM131734).



\section*{APPENDIX A: CONSTRUCT THE SOLUTION LANDSCAPE}

\subsection*{GHiSD algorithm}
The high-index saddle dynamics (HiSD) algorithm aims to find the index-$k$ saddle point($k$-saddle) starting from an initial guess $\boldsymbol u_0$. It is initially designed for gradient systems \cite{yin2020construction}. Given an at least twice differentiable energy function $E(\boldsymbol u)$, where $\boldsymbol u\in \mathbb{R}^n$ is a vector variable, the HiSD algorithm is a pseudo-dynamics depicting the time derivates $\dot{\boldsymbol{u}}$ and $\dot{\\boldsymbol{w}_i}$. Here $\boldsymbol{w}_i$ is the normalized eigenvector corresponding to the $i^{th}$ smallest eigenvalue, where $ i = 1, \cdots k$. The HiSD can be formulated as:
\begin{equation} \label{eqn:HiSD}
\left\{
\begin{aligned}
\dot{\boldsymbol{u}}  &=-\left(\mathbf{I}-\sum\limits_{i=1}^{k} 2\bm{w}_{i} \bm{w}_{i}^{\top}\right) \nabla {E}(\bm{u}),\\
\dot{\bm{w}_i}&=-\left(\mathbf{I}-\boldsymbol{w}_i\boldsymbol{w}_i^\top- \sum\limits_{j=1}^{i-1} 2\bm{w}_j\bm{w}_j^\top\right) G(\bm u)\bm w_i , \\
& \quad i=1, \cdots k.
\end{aligned}
\right.
\end{equation}

where $G(\bm u)$ is the Hessian matrix of $E(\bm{u})$.
A $k$-saddle $\bm u^*$ along with its $k$ unstable eigenvectors $\bm w_1,\cdots, \bm w_k$ will be attained, directly connected by $\bm u_0$. Here $\langle \bm w_i,\bm w_j\rangle=\delta_{ij}$ follows the orthonormal constraint for every t confirmed by Eq.~\ref{eqn:HiSD}.

For the non-gradient system $\dot{\bm{u}}=\bm{F}(\bm{u})$, we modify HiSD by replacing $-G(\bm u)$ with Jacobi operator $J(\bm u)$ and do some changes to the dynamics of $\{ {\bm{w}}_{i} \}$ to guarantee their orthonormality since $J(\bm u)$ is not self-adjoint. Then we get the generalized high-index saddle dynamics (GHiSD):
\begin{equation}\label{eqn:GHiSD}
\left\{
\begin{aligned}
\dot{\bm{u}}=&\left(I-2 \sum_{j=1}^{k} \bm{w}_{j} \bm{w}_{j}^{\top}\right) \bm{F}(\bm{u}), \\
\dot{\bm{w}}_{i}=&\left(I-\bm{w}_{i} \bm{w}_{i}^{\top}\right) J(\bm{u}) \boldsymbol{w}_{i} \\
& - \sum_{j=1}^{i-1} \bm{w}_{j} \bm{w}_{j}^{\top}\left(J(\bm{u})+J^{\top}(\bm{u})\right) \bm{w}_{i}, \quad i=1, \cdots, k.
\end{aligned}
\right.
\end{equation}
In numerical implementation, the above dynamics are discretized in time using explicit Euler schemes on $\bm{u}^{(m)}$ and semi-implicit scheme for $\{ {\bm{w}}_{i} \}$. Since the Jacobian is usually expensive to compute, we adopt the dimer method to avoid evaluating the Jacobian explicitly. 
\begin{equation}\label{eqn:3}
\small
\left\{
\begin{aligned}
&\bm{u}^{(m+1)}=\bm{u}^{(m)}\\
&\qquad \quad +\alpha\left(\bm{F}\left(\bm{u}^{(m)}\right)-2 \sum_{j=1}^{k}\left\langle\bm{F}\left(\bm{u}^{(m)}\right), \bm{w}_{j}^{(m)}\right\rangle \bm{w}_{j}^{(m)}\right) \\
&\tilde{\bm{w}}_{i}^{(m+1)}=\bm{w}_{i}^{(m)}\\
&\qquad \quad +\beta \frac{\bm{F}\left(\bm{u}^{(m+1)}+\bm{l}\bm{w}_{i}^{(m)}\right)-\bm{F}\left(\bm{u}^{(m+1)}-\bm{l} \bm{w}_{i}^{(m)}\right)}{2 l},\\
& \quad i=1, \cdots, k \\
&\left[\bm{w}_{1}^{(m+1)} , \cdots, \bm{w}_{k}^{(m+1)}\right]=\operatorname{orth}\left(\left[\tilde{\bm{w}}_{1}^{(m+1)}, \cdots, \tilde{\bm{w}}_{k}^{(m+1)}\right]\right)
\end{aligned}
\right.    
\end{equation}

where $\alpha$ and $\beta$ are step sizes which are supposed to be sufficently small to ensure the numerical convergence and stability, and $l>0$ is a small constant in the dimer method to give an approximation of the jacobi operator: $J(\bm{u}) \bm{w}_{i} \approx (\bm{F}\left(\bm{u}+\bm{l} \bm{w}_{i}\right)-\bm{F}\left(\bm{u}-\bm{l} \bm{w}_{i}\right))/2 l$.

The initial unstable subspace of $J$, noted as $\mathcal{W}^{\mathrm{u}}(u)$, can be established by iteration of orthonormal vector set $\{\bm w_1,\cdots,\bm w_k\}$ starting from k random initial vectors, which can be regarded as a power method of $I+\beta J$:
\begin{equation}\label{eqn:4}
\small
\left\{
\begin{aligned}
&\tilde{\boldsymbol{w}}_{i}^{(m+1)}=\boldsymbol{w}_{i}^{(m)}+\beta J(\boldsymbol{u}) \boldsymbol{w}_{i}^{(m)}, \quad i=1, \cdots, k \\
&\left[\boldsymbol{w}_{1}^{(m+1)}, \cdots, \boldsymbol{w}_{k}^{(m+1)}\right]=\quad \operatorname{orth}\left(\left[\tilde{\boldsymbol{w}}_{1}^{(m+1)}, \cdots, \tilde{\boldsymbol{w}}_{k}^{(m+1)}\right]\right)
\end{aligned}
\right.    
\end{equation}

In the Schnakenburg model, $\bm{F}$ and $J$ can be formulated as:
\begin{equation}\label{F}
\boldsymbol{F}(\boldsymbol{u}) = \left(\begin{array}{c}
\Delta u +\eta\left(a-u+u^{2} v\right) \\
d \Delta v +\eta\left(b-u^{2} v\right)
\end{array}\right).    
\end{equation}

\begin{equation}\label{J}
J(\boldsymbol{u})\left(\begin{array}{c}
f \\
g
\end{array}\right)=
\left(\begin{array}{c}
\Delta f + \eta (2uv-1)f + \eta u^2 g \\
- 2\eta uv f + d \Delta g  - \eta u^2 g
\end{array}\right),  
\end{equation}
where $\boldsymbol{u}=(u,v)^\top$, $f$ and $g$ are arbitrary $\boldsymbol{C}^2(\Omega) $ functions for spatial variables. 

\subsection*{Downward and upward search}
To identify all stationary states, we combine GHiSD with the following downward and upward search algorithms to construct the solution landscape.

If we pick the initial state $\bm{u}_0$ as a k-saddle and search $(k-1)$-saddles (or lower index saddles) using GHiSD, we are able to construct dynamic connections between saddles: once a new $(k-1)$-saddle is found, an arrow from $\bm{u}_0$ to this saddle is drawn in the solution landscape to illustrate this relationship (see Fig.~\ref{fig2} for illustration). Based on $(k-1)$-saddles we can repeat the process above and construct a map revealing the evolutionary interrelationship between stationary states which we name the solution landscape. We call this kind of strategy searching from top to bottom the downward search strategy \cite{yin2020construction}.

The downward search for $m$-saddles ($m<k$) can be numerically realized as follows. Given a $k$-saddle $u^*$ as the parent state, along with $k$ unstable orthonormal vectors $\{ {\bm{w}}_{i} \}_{i=1}^k$ in $\mathcal{W}^{\mathrm{u}}(u)$, we choose a linear combination of $\{ {\bm{w}}_{i} \}_{i=1}^k$, denoted as $\hat{{\bm{w}}}$, and set the starting state for GHiSD as $u_0=u^* \pm \epsilon \hat{{\bm{w}}}$ to give a small perturbation. Then a $m$-GHiSD starts at $u_0$ with $m$ orthonormal vectors in $\mathcal{W}^{\mathrm{u}}(u)$ perpendicular to $\hat{{\bm{w}}}$. A typical choice for $m$-GHiSD downward search is $\left(\bm{u}^* \pm \epsilon \bm{w}_{m+1}; \bm{w}_{1}, \cdots, \bm{w}_{m}\right)$.

We can also use upward search when the parent state(s) is unknown or we need the information of higher index saddles based on the existing sinks or saddles. The search for $m$-saddles ($m>k$) can be implemented as follows. Choose a stable direction $\hat{{\bm{w}}}$ and $m$ initial directions including $\hat{{\bm{w}}}$. A typical choice for $m$-GHiSD upward search is $\left(\bm{u}^* \pm \epsilon \bm{w}_{m}; \bm{w}_{1}, \cdots, \bm{w}_{m}\right)$. Combining the downward and upward search, we can construct the solution landscape of general dynamic systems using GHiSD.

\section*{APPENDIX B: LINEAR ANALYSIS OF THE SCHNAKENBURG MODEL}
We calculate Morse index of H in the presence of diffusion in the 2D Schnakenburg model Eq.~\ref{Schnakenburg}.

Linear analysis is implemented as follows. Let the small-perturbation of H be:
\begin{equation}\label{eqn:6}
\widetilde{\boldsymbol{u}}(\bm{x},t)=\left(\begin{array}{c}
u-u_{0} \\
v-v_{0}
\end{array}\right).
\end{equation}

Now we look for the solutions of the form:
\begin{equation}\label{eqn:7}
\tilde{\boldsymbol{u}}(\mathbf{r}, t)=\sum\limits_{k} c_{k} \mathbf{W}_{k}(\mathbf{r}) e^{\lambda t},
\end{equation}
where $\mathbf{W}_{k}(\mathbf{r})$ is a time-independent term satisfying a homogeneous Helmholtz equation with non-flux boundary condition:
\begin{equation}\label{eqn:8}
\nabla^{2} \mathbf{W}+k^{2} \mathbf{W}=0.
\end{equation}
The solution of Eq.~\ref{eqn:8} takes the form:
\begin{equation}\label{eqn:9}
\mathbf{W}=\left(\begin{array}{c}
A\cos(m\pi x)\cos(n \pi y) \\
B\cos(m\pi x)\cos(n \pi y)
\end{array}\right).
\end{equation}
Substituting Eq.~\ref{eqn:7}, Eq.~\ref{eqn:9} into Eq.~\ref{eqn:Schnakenburg} and taking $a=\frac{1}{3}, b=\frac{2}{3}$ and $\kappa=k^2=\pi^2 (m^2+n^2)$, we get the secular equation:
\begin{equation}\label{eqn:10}
\left | \begin{array}{cc}
-\kappa+\frac{\eta}{3}-\lambda & \eta \\
-\frac{4}{3} \eta & -\kappa d-\eta-\lambda
\end{array}\right|
=0.
\end{equation}
Thus we have $\lambda_1+\lambda_2=-(\kappa d+\kappa+\frac{2}{3}\eta)<0$, and then a $\lambda$ with positive real part exists if and only if $\lambda_1 \lambda_2=\kappa^2 d+\kappa \eta-\frac{1}{3}\eta \kappa d+\eta^2<0$. Finally we get the condition for Turing instability:
\begin{equation}\label{eqn:11}
d>\min_{\kappa \in K} \frac{\eta(\kappa+\eta)}{\frac{\eta}{3}\kappa-\kappa^2},
\end{equation}
where $K=\{\pi^2 (m^2+n^2)|m,n \in \mathbb{N}\}$. 
Morse index of the H state is obtained by counting the number of possible $(m,n)$ pairs in $K$ with a fixed $d$ satisfying:
\begin{equation}\label{eqn:12}
d>\frac{\eta(\kappa+\eta)}{\frac{\eta}{3}\kappa-\kappa^2},
\end{equation}
By choosing $\eta=200, a=\frac{1}{3}, b=\frac{2}{3}$, Morse index of the H state under different $d$ is listed in Table.~1 of the text.

\section*{APPENDIX C: DIRECT SIMULATION OF THE SCHNAKENBURG MODEL}
Turing pattern can be obtained by using direct simulation of the Schnakenburg model.
When the H state is unstable, we can add small perturbations to H as the initial guess, i.e.,
 $\boldsymbol{u}=\left(u_{0}+\delta u ,v_{0}+\delta v\right)^{\top}$.
Here the perturbations $\delta u$ and $\delta v$ can be generated as follows,
\begin{equation}\label{eqn:13}
\left(\begin{array}{c}
\delta u \\
\delta v
\end{array}\right)
= \sum\limits_{m,n \leq N}\left(\begin{array}{c}
A_{m,n}\cos(m\pi x)\cos(n \pi y) \\
B_{m,n}\cos(m\pi x)\cos(n \pi y)
\end{array}\right),
\end{equation}
where N is a large integer and ${A_{m,n}},{B_{m,n}}$ are uniform distributions sampled from $[-\epsilon/\sqrt{N}, \epsilon/\sqrt{N}]$, where $\epsilon$ is a small positive value to control the intensity of the disturbance.

We numerically solve Eq.~\ref{eqn:Schnakenburg} using the explicit Euler scheme in time and the finite volume scheme in space. 
By taking $\eta=200, d=46$ and different initial $\boldsymbol{u}$, we obtained five sinks T-, D, S, P, C by solving Eq.~\ref{eqn:Schnakenburg} (see Fig.~2 of the text). These results are in consistence with the sinks derived in the solution landscape by using H as the parent state.

\section*{APPENDIX D: ACTION-DEVIATION RELATION}

\subsection*{Definition of action-deviation function}
For the field dynamics defined by PDE of the form: $\dot{\psi}=F(\psi)=\Delta \psi + f(\psi)$,  Lagrangian in Eq.~\ref{actionInt} is given by the Lagrangian density function $\mathcal{L}(\psi(x), \dot{\psi}(x))$:
\begin{equation}
     L(\psi, \dot{\psi})= \int_{\Omega} \mathcal{L}(\psi(x), \dot{\psi}(x)) \mathrm{d} x 
\end{equation}
\begin{equation}
      \mathcal{L}(\psi(x), \dot{\psi}(x)) = \frac{1}{2} \left( \dot{\psi}-\Delta \psi-f(\psi) \right)^2 
\end{equation}
In our 3-species Turing model, $\Omega=(0,6)$. If the translation path $\varphi(s),s \in [0,1]$ corresponding to deviation $\Delta$ ($\Delta \geq 0$ for instance) is parameterized by:
\begin{equation} \label{shift}
\begin{aligned}
&\varphi_{\Delta}(s)(x) \equiv \boldsymbol{u}_{s\Delta}(x) \\
&\equiv \begin{cases}(u_1(x-s\Delta),u_2(x-s\Delta),u_3(x-s\Delta)) & \text {if } s\Delta< x \leq 6 \\
(u_1(0),u_2(0),u_3(0)) & \text {if } 0 \leq x \leq s\Delta \end{cases}
\end{aligned}
\end{equation}
we can establish an explicit formula of action for $\Delta$ shift along this path:
\begin{equation} \label{action1}
\begin{aligned}
S(\Delta) \equiv \hat{S}(\varphi_{\Delta})= & \Delta \int_0^1 \| \boldsymbol{u}_{s\Delta}'(x) \|_{L^2(\Omega)} \cdot \|F(\boldsymbol{u}_{s\Delta}(x))\|_{L^2(\Omega)} \\
& + \left \langle \boldsymbol{u}_{s\Delta}'(x),F(\boldsymbol{u}_{s\Delta}(x))\right \rangle_{L^2(\Omega)} \mathrm{d} s    
\end{aligned}
\end{equation}
where $\left \langle \cdot \right \rangle_{L^2(\Omega)}$ is the inner product under $L^2(\Omega)$ norm. This path $\varphi_{\Delta}(s), s \in [0,1]$ corresponds to a rightward shift process with constant speed $v=\Delta$ connecting the starting pattern $\boldsymbol{u}_0(x) = (u_1(x),u_2(x),u_3(x))$ and the end pattern $\boldsymbol{u}_{\Delta}(x)$. Prolongation of left boundary of $\boldsymbol{u}_{0}(x)$ is introduced artificially in Eq.~\ref{shift} and obeys boundary conditions on the left (if $\Delta \leq 0$, the boundary conditions are satisfied on the right).

\subsection*{Perturbational effect of X3 on the action-deviation function}
For the translation path defined in Eq.~\ref{shift}, the second term of Eq.~\ref{action1} vanishes. $\| \boldsymbol{u}_{s\Delta}'(x) \|_{L^2(\Omega)}$ and $ \|F(\boldsymbol{u}_{s\Delta}(x))\|_{L^2(\Omega)}$ in the integral above can be explicitly expressed by the starting pattern $\boldsymbol{u}_0(x) = (u_1(x),u_2(x),u_3(x))$:
\begin{equation} \label{actionPart}
\begin{aligned}
\| \boldsymbol{u}_{s\Delta}'(x) \|^2_{L^2(\Omega)} =& \int_0^{6-s\Delta} \sum_{i=1}^3 \left( \frac{\mathrm{d} u_i(x)}{\mathrm{d} x}\right)^2 \mathrm{d}x     \\
\|F(\boldsymbol{u}_{s\Delta}(x))\|^2_{L^2(\Omega)} =&  s \Delta \sum_{i=1}^3 
 \left( D_i \frac{\mathrm{d}^2 u_i(x)}{\mathrm{d} x^2}\bigg|_{x=0} \right)^2 \\
 + \Lambda & \sum_{i=1}^3 \left(D_i \frac{\mathrm{d} u_i(x)}{\mathrm{d} x} \bigg|_{x=6-s\Delta-0}-0 \right)^2
\end{aligned}
\end{equation}

Where the last term in the expression of $\|F(\boldsymbol{u}_{s\Delta}(x))\|^2_{L^2(\Omega)}$ is caused by boundary leaping of $\frac{\mathrm{d} u_i(x)}{\mathrm{d} x}$ given non-flux boundary conditions which results in the singularity when calculating the Laplace term of $F$. This can be viewed as the penalty term due to violation of boundary conditions. $\Lambda$ is the penalty factor. Large $\Lambda$ makes the choice of prolongation less important (and we may simply consider the very simple prolongation in Eq.~\ref{shift} ). Actually according to expansion:
\begin{equation} \label{Taylor}
\begin{aligned}
        \frac{\mathrm{d} u_i(x)}{\mathrm{d} x} \bigg|_{x=6-s\Delta} &\approx \frac{\mathrm{d} u_i(x)}{\mathrm{d} x} \bigg|_{x=6}-s\Delta \frac{\mathrm{d}^2 u_i(x)}{\mathrm{d} x^2} \bigg|_{x=6}\\
        &= -s\Delta \frac{\mathrm{d}^2 u_i(x)}{\mathrm{d} x^2} \bigg|_{x=6}
\end{aligned}
\end{equation}
we have 
\begin{small}
\begin{equation}
\begin{aligned}
S(\Delta) &\approx \Delta^2 \sqrt{\Lambda} \left( \sum_{i=1}^3 
\left( D_i \frac{\mathrm{d}^2 u_i(x)}{\mathrm{d} x^2}\bigg|_{x=6} \right)^2 \right)^{\frac{1}{2}} \int_0^1 \| \boldsymbol{u}_{s\Delta}' \|_{L^2(\Omega)}   s \mathrm{d} s \\
 &\approx \frac{1}{2} \Delta^2 \sqrt{\Lambda} \left( \sum_{i=1}^3 
\left( D_i \frac{\mathrm{d}^2 u_i(x)}{\mathrm{d} x^2}\bigg|_{x=6} \right)^2 \right)^{\frac{1}{2}} \| \boldsymbol{u}_{0}' \|_{L^2(\Omega)}
\end{aligned}
\end{equation}
\end{small}

Thus $C$ in Eq.~\ref{C-tau} roughly depends on the second derivative of solution on the boundary:
\begin{equation} \label{C}
    C \approx \frac{1}{2} \sqrt{\Lambda} \left( \sum_{i=1}^3 
\left( D_i \frac{\mathrm{d}^2 u_i(x)}{\mathrm{d} x^2}\bigg|_{x=6} \right)^2 \right)^{\frac{1}{2}} \| \boldsymbol{u}_{0}' \|_{L^2(\Omega)}
\end{equation}

The relationship between $C$ and $\tau$ is shown in the dependence of solution $(u_1(x),u_2(x),u_3(x))$ on $\tau$. By treating $\tau$ as a small coefficient we could calculate the first order perturbation of solutions based on Eq.~\ref{3-species}. Let
\begin{equation} \label{expansion}
\begin{aligned}
        u_i(x;\tau)=u_i^0(x)+\tau u_i^1(x)+\tau^2 &u_i^2(x)+\cdots\\
    &i=1,2,3
\end{aligned}
\end{equation}
and $u_i^0(x),i=1,2$ is the 2-species solution in the absence of $X_3$. $u_3^0(x)$ is a constant. The $ \mathcal O(\tau)$ term of Eq.~\ref{3-species} gives the equation to compute first order perturbation function $u_i^1(x),i=1,2,3$. This equation is a linear PDE:
\begin{equation} \label{linear approx}
    \bm{D} \frac{\mathrm{d}^2 \bm{u^1}(x)}{\mathrm{d} x^2} =\bm{A}(x)\bm{u^1}(x)+\bm{c}(x)
\end{equation}
where $\bm{u^1}(x)=(u_1^1(x),u_2^1(x),u_3^1(x))^{\top}$ and satisfies non-flux boundary conditions. $\bm{D}$ is the diffusion coefficient matrix. $\bm{A}=\bm{A}(x)$ and $\bm{c}=\bm{c}(x)$ are:

 \begin{equation}
 \begin{small}
  \bm{D}=\left[
 \begin{array}{ccc}
     D_1 & 0 & 0 \\
     0 & D_2 & 0 \\
     0 & 0 & D_3 
 \end{array}
 \right]         
 \end{small}      
 \end{equation}

 \begin{equation}
 \begin{small}
  \bm{A}=\left[
 \begin{array}{ccc}
     \tilde{k}_{12}(u_2^0)^2+k_{12} & 2\tilde{k}_{12}u_2^0 u_1^0 -3\tilde{k}_{21}(u_2^0)^2 - k_{21} & 0 \\
     -(\tilde{k}_{12}(u_2^0)^2+k_{12}) & -(2\tilde{k}_{12}u_2^0 u_1^0 -3\tilde{k}_{21}(u_2^0)^2 - k_{21}) & 0 \\
     0 & 0 & 0 
 \end{array}
 \right]         
 \end{small}      
 \end{equation}

\begin{equation}
  \bm{c}=\left[
 \begin{array}{c}
     \frac{k_{13}}{\tau}u_1^0 - \frac{k_{31}}{\tau}u_3^0 \\
 \frac{k_{23}}{\tau}u_2^0-\frac{k_{32}}{\tau}u_3^0 \\
     - \left(\frac{k_{13}}{\tau}u_1^0 - \frac{k_{31}}{\tau}u_3^0 \right) - \left(\frac{k_{23}}{\tau}u_2^0-\frac{k_{32}}{\tau}u_3^0 \right)
 \end{array}
 \right]             
 \end{equation}

According to Eq.~\ref{expansion}, Eq.~\ref{C} can be formulated in the expansion of $\tau$:
\begin{equation}
\begin{small}
        C(\tau)=\frac{\sqrt{\Lambda}}{2} M(\bm{u}^0) 
      \times \left( 1+\tau R(\bm{u}^0,\bm{u}^1) + \mathcal{O}(\tau^2)\right)  
\end{small}
\end{equation}

where 
\begin{equation}\label{MR}
\begin{small}
    \begin{aligned}
        M(\bm{u}^0)=&\left(  \sum\limits_{i=1}^3 
        \left( D_i \frac{\mathrm{d}^2 u_i^0(x)}{\mathrm{d} x^2}\bigg|_{x=6} \right)^2 \right)^{\frac{1}{2}} \left(  \sum\limits_{i=1}^3 \int_0^6 \left( \frac{\mathrm{d} u_i^0}{\mathrm{d} x} \right)^2 \mathrm{d} x
          \right)^{\frac{1}{2}} \\ 
        R(\bm{u}^0,\bm{u}^1)&=\frac{\sum\limits_{i=1}^3 
         D_i^2 \frac{\mathrm{d}^2 u_i^0(x)}{\mathrm{d} x^2}\big|_{x=6} \frac{\mathrm{d}^2 u_i^1(x)}{\mathrm{d} x^2}\big|_{x=6}}{\sum\limits_{i=1}^3 
        \left( D_i \frac{\mathrm{d}^2 u_i^0(x)}{\mathrm{d} x^2}\big|_{x=6} \right)^2}+\frac{\sum\limits_{i=1}^3 \int_0^6  \frac{\mathrm{d} u_i^0}{\mathrm{d} x}\frac{\mathrm{d} u_i^1}{\mathrm{d} x}  \mathrm{d} x}{\sum\limits_{i=1}^3 \int_0^6 \left( \frac{\mathrm{d} u_i^0}{\mathrm{d} x} \right)^2 \mathrm{d} x} \\
        &=\frac{\sum\limits_{i=1}^2 
         D_i^2 \frac{\mathrm{d}^2 u_i^0(x)}{\mathrm{d} x^2}\big|_{x=6} \frac{\mathrm{d}^2 u_i^1(x)}{\mathrm{d} x^2}\big|_{x=6}}{\sum\limits_{i=1}^2 
        \left( D_i \frac{\mathrm{d}^2 u_i^0(x)}{\mathrm{d} x^2}\big|_{x=6} \right)^2}+\frac{\sum\limits_{i=1}^2 \int_0^6  u_i^0 \frac{\mathrm{d}^2 u_i^1(x)}{\mathrm{d} x^2}  \mathrm{d} x}{\sum\limits_{i=1}^2 \int_0^6 u_i^0 \frac{\mathrm{d}^2 u_i^0(x)}{\mathrm{d} x^2} \mathrm{d} x}
    \end{aligned}
\end{small}
\end{equation}

The last step uses integration by parts and $u_3^0(x)=const.$ . 
We get the final estimation for $\epsilon$ and $\eta$ defined in Eq.~\ref{expan}:
\begin{equation}
    \begin{aligned}
\epsilon&= \frac{\sqrt{\Lambda}}{2} M(\bm{u}^0) \\
\eta    &= \frac{\sqrt{\Lambda}}{2} M(\bm{u}^0)R(\bm{u}^0,\bm{u}^1)
    \end{aligned}
\end{equation}

The ratio $\dfrac{\eta}{\epsilon}$ is exactly the linear coefficient $R(\bm{u}^0,\bm{u}^1)$. $R(\bm{u}^0,\bm{u}^1) \gg 1$ can be simply understood through the fact that:
\begin{equation} \label{eps_eta}
\begin{aligned}
\left| \frac{\mathrm{d}^2 u_i^0(x)}{\mathrm{d} x^2}\big|_{x=6} \right| \ll  \left| \frac{\mathrm{d}^2 u_i^1(x)}{\mathrm{d} x^2}\big|_{x=6} \right| ,& i=1,2 \\
 \frac{\mathrm{d}^2 u_i^0(x)}{\mathrm{d} x^2}\big|_{x=6} \times \frac{\mathrm{d}^2 u_i^1(x)}{\mathrm{d} x^2}\big|_{x=6}>0   , & i=1,2. \\
\end{aligned}
\end{equation}

Actually small second derivatives of $\bm{u}^0$ stem from the almost free shift in two-species case. For example when $\bm{u}$ is the solution S1, $D_1 \frac{\mathrm{d}^2 u_1^0(x)}{\mathrm{d} x^2}\big|_{x=6} \approx D_2 \frac{\mathrm{d}^2 u_2^0(x)}{\mathrm{d} x^2}\big|_{x=6} \approx 10^{-10}$, but $u_i^1(x)$ is regularly solved from linear equation Eq.~\ref{linear approx} thus
$\frac{\mathrm{d}^2 u_i^1(x)}{\mathrm{d} x^2}\big|_{x=6} \sim \mathcal O(1), i=1,2$. 
From Eq.~\ref{MR} roughly we know $M(\bm{u}^0) \propto \sum\limits_{i=1}^3 D_i \frac{\mathrm{d}^2 u_i^0(x)}{\mathrm{d} x^2}\bigg|_{x=6} $ and $R(\bm{u}^0,\bm{u}^1) \propto \left( \sum\limits_{i=1}^3 D_i \frac{\mathrm{d}^2 u_i^0(x)}{\mathrm{d} x^2}\bigg|_{x=6} \right)^{-1}$.
In this sense, $\epsilon$ depends linearly on the second derivatives of species' concentration on the boundary, which vary widely from solution to solution and could be very small. But $\eta$ is more or less the same for different sinks.

\bibliographystyle{naturemag}
\bibliography{ref}

\end{document}


\maketitle


\subsection*{Attraction basins of five sinks in Schnakenburg model in subcritical regime}
\begin{figure*}
\centering
\includegraphics[width=\linewidth]{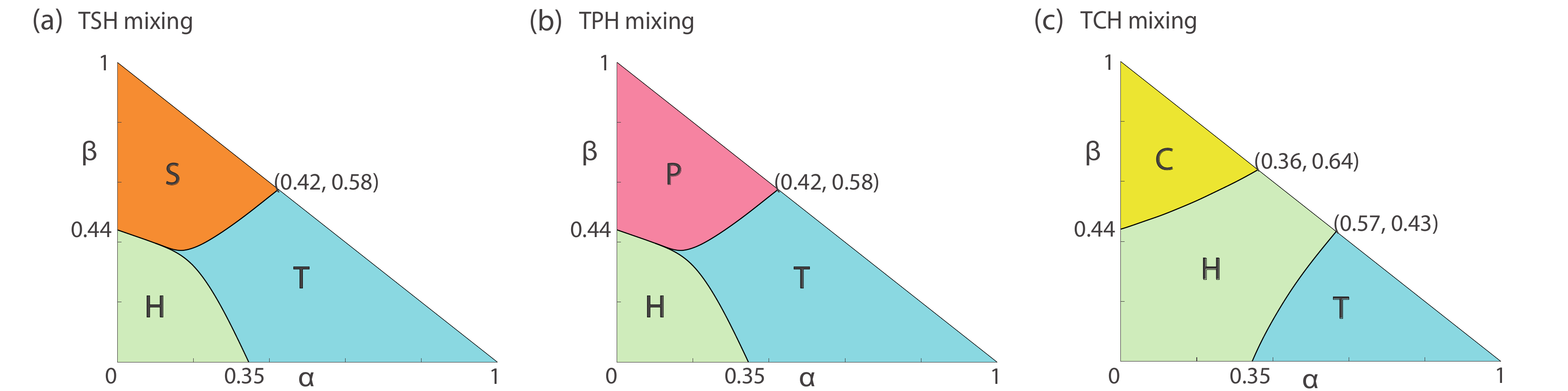}
\caption{The attraction basins of five sinks in Schnakenburg model at $d=39$. The initial states are chosen to be convex combination expressed as $\alpha $T$+\beta$X$+(1-\alpha -\beta)$H, where X=S (a), P (b), C (c). (a) and (b) are exactly the same, and the attraction basin is separated into three regions in the domain $D=\{ (\alpha, \beta) | \alpha \geq 0, \beta \geq 0 , \alpha + \beta \leq 1  \}$ by three critical lines. (c) the attraction basins of C and T are separated by H.}
\label{smfig1}
\end{figure*}

Furthermore, we exam the attraction basins for five sinks (i.e., T, S, P, C, H) at $d=39$ through the numerical simulations of Eq.3 in the text by choosing different convex combination of five sinks as the initial states and evolving to steady states asymptotically.
Fig.~\ref{smfig1} shows three experiments mixing T,S,H sinks, T,P,H sinks and T,C,H sinks respectively. They all indicate that T owns the largest basin of attraction( in Fig.3 in the text, T is shown to be the first to emerge). Additionally these four inhomogeneous sinks have quite large basins of attraction, indicating that these sinks are non-negligible subcritical patterns before the onset of the Turing instability.

\subsection*{The influence of boundary conditions}
\begin{figure*}[bth]
\centering
\includegraphics[width=12cm]{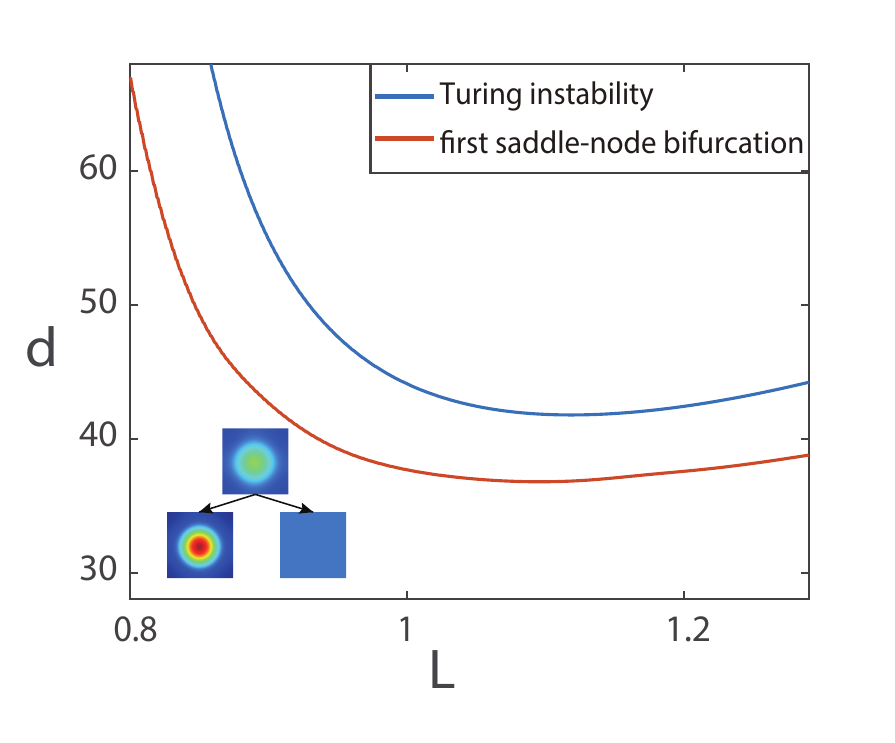}
\caption{Phase diagram for different $d$ and $L$ with periodic boundary condition ($\eta=204$).}
\label{smfig2}
\end{figure*}

To test the influence of boundary conditions on Turing pattern, we change no-flux boundary condition to periodic boundary condition in the Schnakenburg model. The corresponding phase diagram shows that the first saddle-node bifurcation always occurs before Turing instability (Fig.~\ref{smfig2}), which is consistent with the phase diagram with no-flux boundary condition. But the first inhomogenous sink to emerge via saddle-node bifurcation is always the same because of the constraints of periodic boundary conditions to the possible solutions.

\subsection*{Pattern formation in Gierer-Meinhardt model}

To verify the universal nature of nonlinear pattern-formation mechanism beyond Turing mechanism, we further examine the Gierer-Meinhardt model \cite{gierer1972theory, RevModPhys.66.1481}, which is a classical RD model of the activator-inhibitor type that accounts for many important types of pattern formation and morphogenesis such as shells of molluscs \cite{meinhardt1987model}, regulation in the Stripe-like patterns of zebrafish \cite{zebrafish}, and formation of net-like structures of differentiated cells \cite{meinhardt2009algorithmic}. We investigate the following Gierer-Meinhardt model in a square domain: 
\begin{equation}
\begin{aligned}
&\frac{\partial u}{\partial t}=a-b u+\frac{u^2}{v\left(1+c u^2\right)}+D_u \Delta u ,\\
&\frac{\partial v}{\partial t}=u^2-v+D_v \Delta v .
\end{aligned}    
\end{equation}

Here $u$ and $v$ are the concentrations of short-range activator and long-range inhibitor respectively. Parameters $a, b, c$ depict the chemical reactions and $D_u, D_v$ denote the spatial diffusion rates. 
The boundary condition of the Gierer-Meinhardt model is set as the no-flux boundary condition. We choose two different domain sizes of the system:  $1 \times 1$ square and $5 \times 5$ square. The parameters are given by $a=0, b=1.2, c=0.4, D_u=1 \times 10^{-2},$ and $D_v= 11 \times 10^{-2}$ in the numerical simulations.

\begin{figure}[htb]
\centering
\includegraphics[width=9cm]{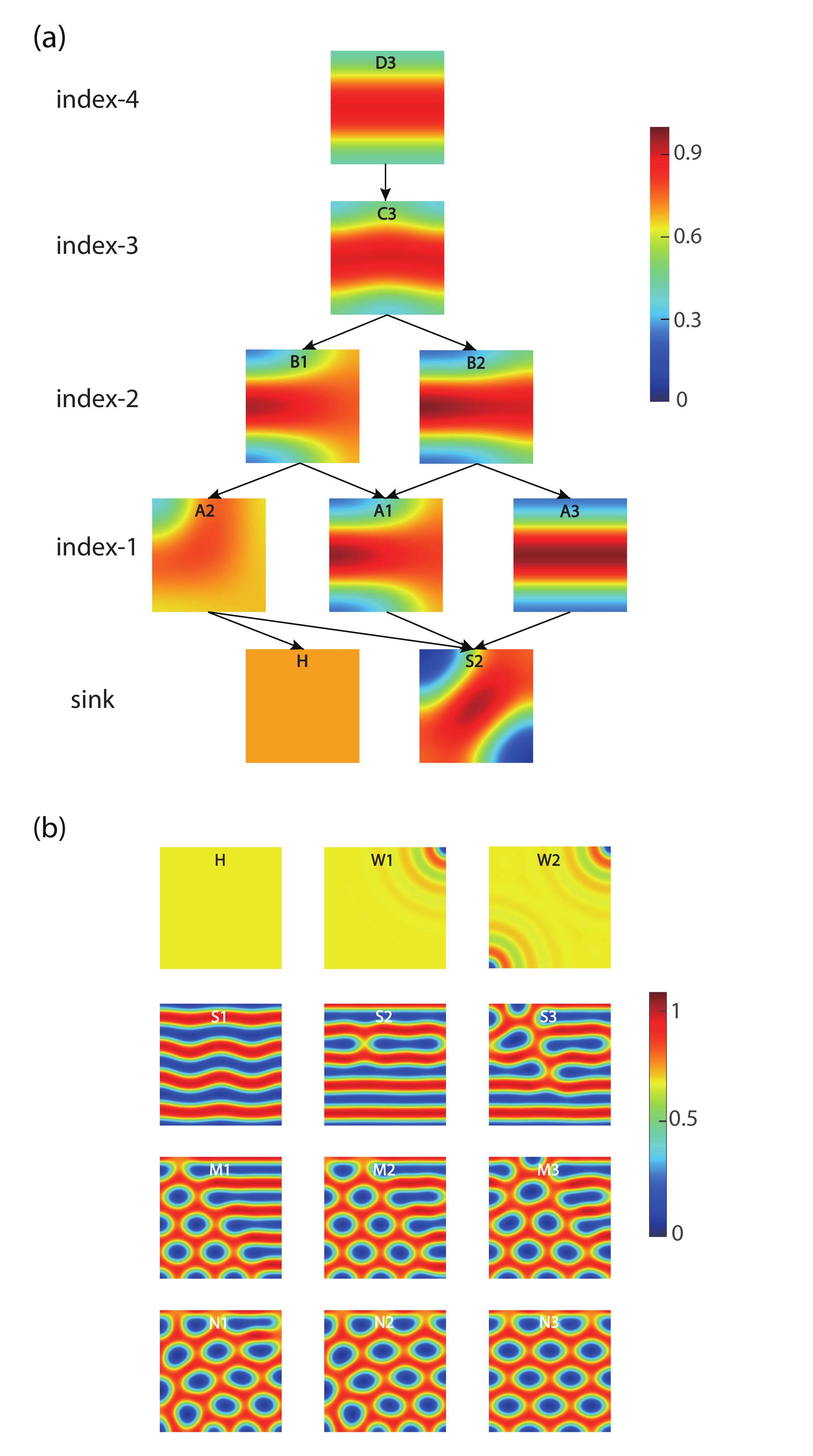}
\caption{(a) Solution landscape of the Gierer-Meinhardt model in the $1\times1$ small square. (b) Some typical stable states in the solution landscape in the $5\times5$ square. Patterns of multiple shapes emerge when the system size becomes larger.
}\label{smfig3}
\end{figure}

We first construct the solution landscape of the Gierer-Meinhardt model in a small $1\times 1$ square in Fig.~\ref{smfig3}(a). Because the reactants are strongly influenced by the domain size and boundary, all spatial heterogeneous stationary states are the stripe patterns in a small square due to the dispersion relation.
The stripe state D3, when we account for a rotational symmetry, is the only highest-index (index-4) saddle point in the solution landscape. Other stationary states can be found using the downward search algorithm.
This solution landscape shows that, while the H state remains a sink, a new stable sink S2 with lower symmetry emerges, indicating that the Turing mechanism is not the prerequisite to generate spatial patterns in the Gierer-Meinhardt model.

The spatial patterns are much more abundant in the large $5 \times 5$ square. By the construction of solution landscape under the same parameters as in $1 \times 1$ case, multiple stable sinks can be found by the GHiSD method. As an illustration (Fig.~\ref{smfig3}(b)), we show some typical stable states, including stripe patterns (S1-S3), net-like patterns (N1-N3) and wave-like patterns (W1 and W2). M1-M3 can be regarded as the mixture of stripe and net-like patterns. These spatial patterns also co-exist with the H state, which again demonstrate the existence of a generic nonlinear mechanism that is beyond the scope of Turing mechanism.



\clearpage
\subsection*{Bifurcation diagram and eigenvalues of solutions in 3-species model}

\begin{figure}[htb]
    \centering
    \includegraphics[width=\linewidth]{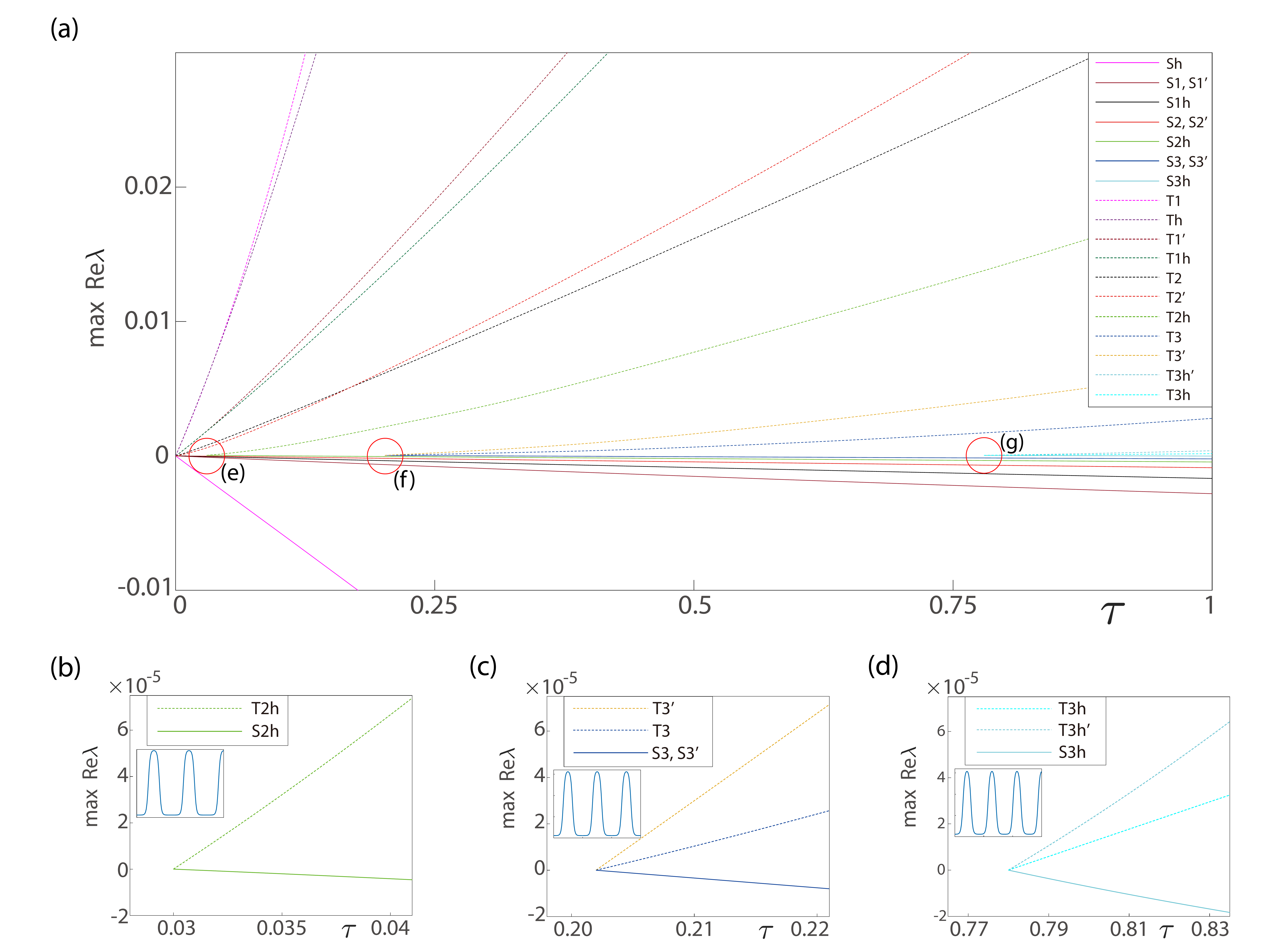}
    \caption{(a) The maximum value of the real part of eigenvalues of sinks and saddles (denoted by $\lambda$) under $\tau \in [0,1]$ . Dotted lines stand for 1-saddles and solid lines stand for sinks. Some sinks and saddles already exist at $\tau=0$ while others form at positive $\tau$.  
(b)-(d) Later-formed sinks and saddles through bifurcation. }
    \label{smfig4}
\end{figure}

\clearpage
\subsection*{Action-deviation relation at different $\tau$}

\begin{figure}[htb]
    \centering
    \includegraphics[width=\linewidth]{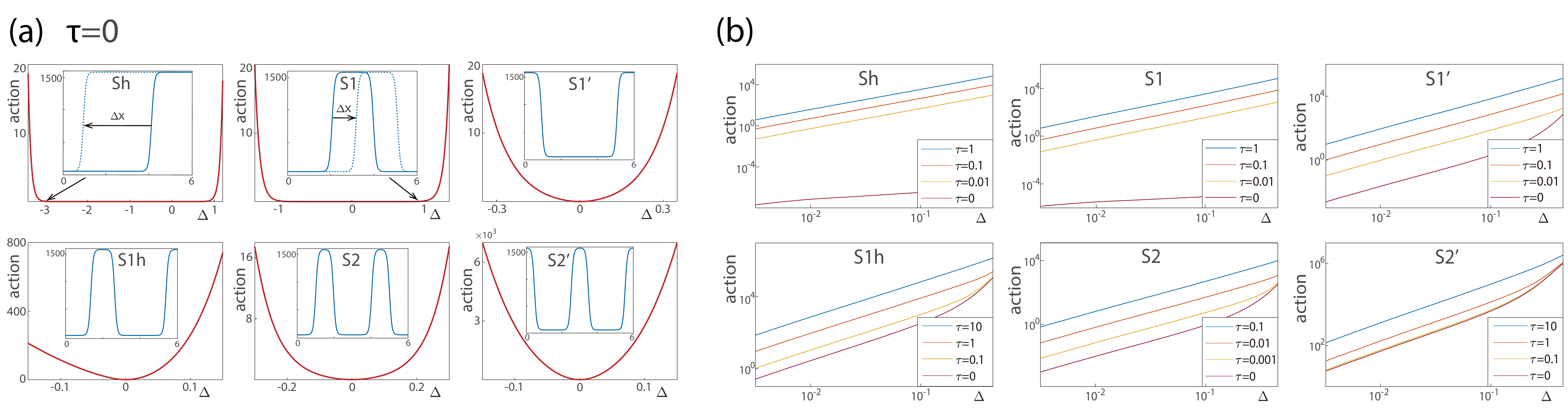}

    \caption{ 
(a) Action-deviation ($S-\Delta$) relations (red lines) for 6 sinks which have already appeared in the 2-specie model ($\tau=0$). The solution profiles for  specie-2: $u_2(x)$ are shown in the insets. Note that the action-deviation curves for Sh and S1 sinks are flat near $\Delta=0$ reflecting the approximate translational symmetry of the two sinks as shown in the corresponding insets. 
 (b)Action-deviation relations of the six sinks in (a) for different values of $\tau$. The slope $\approx 2$ for every sink shows the quadratic relation. When $\tau$ is far from 0, the uniformly-spaced curves in each graph of Fig.5(b) indicate the validity of the linear expansion in Eq.8 in the text.  
 }
    \label{smfig5}
\end{figure}

\clearpage
\bibliographystyle{naturemag}
\bibliography{ref}